\documentclass[twocolumn]{autart}    
\usepackage{amsmath,amssymb}
\usepackage{bm}

\newtheorem{hyp}{\bf Assumption}

\usepackage{tikz}
\usetikzlibrary{arrows,shapes,calc}
\usepackage{pgfplots}
\usetikzlibrary{positioning} 

\tikzstyle{solid}=                   [dash pattern=]
\tikzstyle{dotted}=                  [dash pattern=on \pgflinewidth off 2pt]
\tikzstyle{densely dotted}=          [dash pattern=on \pgflinewidth off 1pt]
\tikzstyle{loosely dotted}=          [dash pattern=on \pgflinewidth off 4pt]
\tikzstyle{dashed}=                  [dash pattern=on 3pt off 3pt]
\tikzstyle{densely dashed}=          [dash pattern=on 3pt off 2pt]
\tikzstyle{loosely dashed}=          [dash pattern=on 3pt off 6pt]
\tikzstyle{dashdotted}=              [dash pattern=on 3pt off 2pt on \the\pgflinewidth off 2pt]
\tikzstyle{densely dashdotted}=      [dash pattern=on 3pt off 1pt on \the\pgflinewidth off 1pt]
\tikzstyle{loosely dashdotted}=      [dash pattern=on 3pt off 4pt on \the\pgflinewidth off 4pt]

\usepackage{graphicx}      
\begin{document}

\begin{frontmatter}

\title{Stability proof for nonlinear MPC design using monotonically increasing weighting profiles without terminal constraints} 


\author[Mazen]{Mazen Alamir}\ead{mazen.alamir@grenoble-inp.fr},    

\address[Mazen]{CNRS, Univ. Grenoble Alpes, Gipsa-lab, F-3800, Grenoble, France.}  


\begin{abstract}                          
In this note, a new formulation of Model Predictive Control (MPC) framework with no stability-related terminal constraint is proposed and its stability is proved under mild standard assumptions. The novelty in the formulation lies in the use of time-varying monotonically increasing stage cost penalty. The main result is that the $0$-reachability  prediction horizon can always be made stabilizing provided that the increasing rate of the penalty is made sufficiently high. 
\end{abstract}

\end{frontmatter}

\section{Introduction}
In the majority of NMPC formulations, the stability of the closed-loop implies the use of terminal constraints on the state. In the early formulations \cite{Keerthi1988,Mayne1990,alamirscl94}, a strong point-wise equality constraint on the state was introduced. This constraint is imposed at the end of the prediction horizon, namely $N$-steps ahead where $N$ is such that the targeted state is $N$-step reachable. 

Since, relaxations were introduced through the combined use of terminal set constraint and appropriate terminal penalty. The many different ways to choose these two items were unified in \cite{Mayne2000} where it has been shown that the terminal set should be controlled-invariant under some {\em local}  feedback control that makes the terminal penalty a control-Lyapunov function inside the terminal set. The difficulty in computing the terminal set and the associated Lyapunov function for general nonlinear systems remains quite dissuasive in real-life applications. 

On the other hand, it has been shown quite early \cite{Alamir1995} that provable stability can be obtained without terminal stability-related constraint by using {\em sufficiently long} prediction horizon \cite{Grim:2005,Jadbabaie:2005}. More recent results followed, [see \cite{Grune2010,Grune2011,Boccia:2014} and the references therein] where deeper analysis is obtained regarding this fact. However, the underlying argument remained that with sufficiently long prediction horizon, the optimal decisions necessarily lead to open-loop trajectories with terminal appropriate properties. 

Another family of stabilizing formulations without terminal constraint are those based on the contraction property \cite{Kothare2000,alamir2007low,alamir2017}. These formulations are quite attractive in terms of the minimal necessary prediction horizon length although they sometimes involve rather non conventional implementation forms and/or Lyapunov functions. 

This note proposes a flash-back towards the early $N$-reachability related formulations with the exception that no stability-related terminal constraint is used. Instead, a non uniform (in time) penalty is used in the definition of the stage cost. This obviously strengthens the weight on the tail of the prediction horizon leading to similar effects as those induced by the infinite (or sufficiently long) horizon costs. The idea might seem straightforward and it is probably so. Nevertheless, this technical note gives the formal proof of the stability result. To say it shortly, the proposed formulation inherits the combined advantages of early formulations (finite, short and stabilizing horizon) and the infinite horizon formulations (absence of final stability-related constraint). 

The paper is organized as follows: The problem is stated in Section \ref{secdefnot}. The necessary assumptions are given in Section \ref{reqass} before the main result is given in Section \ref{secmainres}. 

It is worth underlying that while only control-related constraints are included in the formulation of the present note, the result of this note can easily be extended to the case where state constraints are present. This would, however, add useless (and quite standard) technicalities that may hide the main underlying arguments. Remark \ref{sconstraints} gives however the main changes that need to be incorporated to handle state constraints.  

\section{Problem statement} \label{secdefnot} 
Let us consider general nonlinear systems of the form:
\begin{equation}
x_{k+1}=f(x_k,u_k) \label{syst} 
\end{equation} 
where $x_k\in \mathbb{R}^{n}$ and $u_k\in \mathbb{R}^{n_u}$ represent the state and the control vectors respectively at instant $k$. It is assumed that $f(0,0)=0$ and that the control objective is to stabilize the steady state $x=0$.
 
Consider control profiles $\bm u:=(u_0,u_1,\dots,u_{N-1})$ defined over a prediction horizon of length $N$. Denote the corresponding state trajectory starting from $x_k$ by $\bm x_i^{\bm u}(x_k)$ for $i=0,\dots,N$, namely:
\begin{align}
\bm x_0^{\bm u}(x_k)&=x_k \label{rec1} \\
\bm x_i^{\bm u}(x_k)&=f(\bm x_{i-1}^{\bm u}(x_k),u_{i-1})\quad \mbox{\rm for}\   i\in \{1,\dots,N\}
\end{align} 
Based on the trajectories $\bm u$ and $\bm x^{\bm u}(x_k)$, let us consider a cost function of the form:
\begin{equation}
J_m(\bm u\vert x_k):=\sum_{i=1}^N\left(i/N\right)^m\ell(\bm x^{\bm u}_i(x_k)) \label{cost} 
\end{equation} 
for some integer $m\in \mathbb N$. This cost enables the following optimization problem to be defined for any compact set $\mathbb U\subset \mathbb{R}^{Nn_u}$ of admissible control profiles:
\begin{equation}
\mathcal P_m(x_k):\quad \min_{\bm u\in \mathbb U} J_m(\bm u\vert x_k) \label{optimprob} 
\end{equation}  
Let us denote a solution to $\mathcal P(x_k)$ (if any) by $\bm u^*(x_k,m)$ and the corresponding optimal cost $J_m^*(x_k)$.

The aim of this note is to investigate the conditions under which $x=0$ is an asymptotically stable equilibrium for the closed-loop system given by:
\begin{equation}
x_{k+1}=f(x_k,\bm u^*(x_k,m)) \label{clsyst} 
\end{equation} 
\subsection{Notation}
In what follows, the subset $B_\ell(\rho)\subset \mathbb{R}^{n}$ denotes the $\rho$-level set of $\ell$, namely $B_\ell(\rho):=\{x\in \mathbb{R}^{n}\  \vert\  \ell(x)\le \rho\}$. 
\section{Statement of the required assumptions} \label{reqass}
The first assumption is a standard $N$-step reachability condition of the targeted state $x=0$. This Assumption is  commonly used in the first provably-stable formulations \cite{Keerthi1988,Mayne1990,alamirscl94}:
\begin{hyp}\label{Ncontr} 
The maps $f$ and $\ell$ are continuous and $\ell$ is positive definite. Moreover, there is a compact set $\mathbb X_N$ such that for all $x\in \mathbb X_N$, the set defined by:
\begin{equation}
\mathbb U_{x\rightarrow 0}:=\left\{\bm u\in \mathbb U\quad \mbox{\rm s.t} \quad  \bm x_N^{\bm u}(x)=0\right\}
\end{equation} 
is not empty.
\end{hyp}
The second assumption is a local control-invariance property that is assumed in the neighborhood of the origin.  
\begin{hyp}\label{voisinage} 
There exists $\bar\rho>0$ such that
\begin{align}
\forall \rho\le \bar\rho, \forall x\in B_\ell(\rho), &\exists u^+\ \mbox{\rm s.t}, \nonumber \\
&\ell(f(x,u^+))-\ell(x)\le -q(x) \label{decrease} 
\end{align} 
for some positive definite function $q$ satisfying:
\begin{equation}
q(x)\ge  \gamma \ell(x) \label{defdegamma} 
\end{equation} 
for some $\gamma>0$ and for all $x\in \mathbb X_N$.
\end{hyp}

This is again a rather standard assumption since (\ref{decrease}) simply means that $\ell$ is a {\bf local} control-Lyapunov function with decrease rate described by $q$. Moreover, the inequality  (\ref{defdegamma}) generically holds for a large choice of $\ell$ and $q$ including positive definite quadratic forms. More precisely, if $\ell(x)=x^TQ_\ell x$ and $q=x^TQ_qx$ then $\gamma:= \frac{\lambda_{min}(Q_q)}{\lambda_{max}(Q_\ell)})$ satisfies the condition (\ref{defdegamma}). 
\section{Main results} \label{secmainres} 
\begin{lem} \label{lem1} 
Under Assumption \ref{Ncontr}, for all $x\in \mathbb X_N$, one has:
\begin{equation}
\ell(\bm x^{\bm u^*(x,m)}_N(x))\le \eta\cdot c^m \quad \mbox{\rm where $c:=\frac{N-1}{N}<1$} 
\end{equation} 
fo some bounded $\eta>0$.
\end{lem}
{\sc Proof.}  Take $\bm u^0\in \mathbb U_{x\rightarrow 0}$ which is possible (since $x\in \mathbb X_N$) by virtue of Assumption \ref{Ncontr}. By the definition of optimality one has:
\begin{align*}
J^*_m(x)&\le J_m(\bm u^0\vert x)\\
&\le \sum_{i=1}^N\left(i/N\right)^m\ell(\bm x^{\bm u^0}_i(x))
\end{align*} 
and since $\ell(\bm x^{\bm u^0}_N(x))=0$ by assumption, the last term can be removed to get:
\begin{align*}
J^*_m(x)&\le \sum_{i=1}^{N-1}\left(i/N\right)^m\ell(\bm x^{\bm u^0}_i(x)) \\
&\le \bigl(\dfrac{N-1}{N}\bigr)^m \sum_{i=1}^{N-1}\ell(\bm x^{\bm u^0}_i(x))
\end{align*}  
This obviously gives the result if  $\eta$ is defined s.t:
\begin{equation}
\eta\ge \sup_{(x,\bm u)\in \mathbb X_N\times \mathbb U^N} \sum_{i=1}^{N-1}\ell(\bm x_i^{\bm u}(x))
\end{equation} 
which is obviously well defined and bounded by continuity of $f$, $\ell$ and the fact that $\mathbb X_N\times \mathbb U^N$ is a compact set. 
$\hfill \Box$  
\begin{prop}\label{prop} 
Under Assumptions \ref{Ncontr} and \ref{voisinage}, the targeted state $x=0$ is asymptotically stable for the closed-loop dynamics (\ref{clsyst}) for all initial state $x\in \mathbb X_N$.  
\end{prop}
{\sc Proof.} Let us shortly denote the optimal profile at instant $k$ by $\bm u^*:=\bm u^*(x_k,m)$. At instant $k+1$, consider the candidate control profil $\tilde{\bm u}_k$ defined by:
\begin{equation}
\tilde{\bm u}:=\left(\bm u_1^*,\dots,\bm u_{N-1}^*,\bar u\right)
\end{equation} 
where $\bar u\in \mathbb U$ is defined by:
\begin{equation}
\bar u:= \mbox{\rm arg}\min_{u\in \mathbb U}\quad \ell\Bigl(f(\bm x^{\bm u^*}_{N}(x_k),u)\Bigr)
\end{equation} 
As a candidate solution to $\mathcal P_m(x_{k+1})$, $\tilde{\bm u}$ corresponds to a cost function satisfying:
\begin{align*}
&J_m(\tilde{\bm u}\vert x_{k+1})=\ell\left(f(\bm x^{\bm u^*}_{N}(x_k),\bar u)\right)\\
&+\sum_{i=1}^{N-1}(i/N)^m\ell(\bm x^{\bm u^*}_{i+1}(x_k))
\end{align*} 
which can be rewritten using the change of indices $j=i+1$ as follows:
\begin{align}
&J_m(\tilde{\bm u}\vert x_{k+1})=\ell\left(f(\bm x^{\bm u^*}_{N}(x_k),\bar u)\right)\nonumber \\
&+\sum_{j=2}^{N}(\frac{j-1}{N})^m\ell(\bm x^{\bm u^*}_{j}(x_k)) \label{hg76} 
\end{align} 
Now for the sake of readability, let us use the following compact notation:
\begin{equation}
\ell^*_j(x_k):=\ell(\bm x^{\bm u^*}_{j}(x_k))
\end{equation} 
with this notation, equation (\ref{hg76}) becomes:
\begin{align}
&J_m(\tilde{\bm u}\vert x_{k+1})=\ell\left(f(\bm x^{\bm u^*}_{N}(x_k),\bar u)\right)\nonumber \\
&+\sum_{j=2}^{N}(\frac{j-1}{N})^m\ell^*_j(x_k) \label{hg76bis} 
\end{align} 
using straightforward neutral operations, it comes that:
\begin{align}
&J_m(\tilde{\bm u}\vert x_{k+1})=\ell\left(f(\bm x^{\bm u^*}_{N}(x_k),\bar u)\right)\nonumber \\
&+\sum_{j=2}^{N}(\frac{j-1}{j})^m(\frac{j}{N})^m\ell^*_j(x_k)\nonumber
\end{align} 
and by adding and removing the same terms:
\begin{align}
&J_m(\tilde{\bm u}\vert x_{k+1})=\ell\left(f(\bm x^{\bm u^*}_{N}(x_k),\bar u)\right)\nonumber \\
&+\sum_{j=2}^{N}\Bigl[(\frac{j-1}{j})^m-1\Bigr](\frac{j}{N})^m\ell^*_j(x_k)\nonumber \\
&+\sum_{j=2}^{N}(\frac{j}{N})^m\ell^*_j(x_k) \label{gfft54} 
\end{align} 
But note that the last term in (\ref{gfft54}) satisfies:
\begin{equation}
\sum_{j=2}^{N}(\frac{j}{N})^m\ell^*_j(x_k)=J^*_m(x_k)-\dfrac{1}{N^m}\ell^*_1(x_k)
\end{equation} 
Using this last equation in (\ref{gfft54}) gives:
\begin{align}
&J_m(\tilde{\bm u}\vert x_{k+1})=J^*_m(x_k)-\dfrac{1}{N^m}\ell^*_1(x_k)+\nonumber \\
&-\sum_{j=2}^{N}\Bigl[1-(\frac{j-1}{j})^m\Bigr](\frac{j}{N})^m\ell^*_j(x_k) \nonumber \\
&+\ell\left(f(\bm x^{\bm u^*}_{N}(x_k),\bar u)\right) \label{ououf} 
\end{align} 
and since for all $j\in \{2,\dots,N\}$, one has:
\begin{equation}
\Bigl[1-(\frac{j-1}{j})^m\Bigr]\ge \Bigl[1-(\frac{N-1}{N})^m\Bigr]=:\psi(m)
\end{equation} 
the equation (\ref{ououf}) implies:
 \begin{align}
&J_m(\tilde{\bm u}\vert x_{k+1})\le J^*_m(x_k)-\dfrac{1}{N^m}\ell^*_1(x_k)-\nonumber \\
&-\psi(m)\sum_{j=2}^{N}(\frac{j}{N})^m\ell^*_j(x_k)+\ell\left(f(\bm x^{\bm u^*}_{N}(x_k),\bar u)\right) \label{ououfbis} 
\end{align} 
and keeping only the last term of the sum in the first term of the second line, one obtains:
 \begin{align}
&J_m(\tilde{\bm u}\vert x_{k+1})\le J^*_m(x_k)-\dfrac{1}{N^m}\ell^*_1(x_k)+\nonumber \\
&-\psi(m)\ell^*_N(x_k)+\ell\left(f(\bm x^{\bm u^*}_{N}(x_k),\bar u)\right) \label{ououfbis} 
\end{align} 
Now according to Lemma \ref{lem1}, $\bm x^{\bm u^*}_{N}(x_k)\in B_\ell(\eta c^m)$ which, together with Assumption \ref{voisinage} implies that for sufficiently high $m$, one has:
\begin{equation}
\eta c^m\le \bar\rho 
\end{equation} 
where $\bar\rho$ is the positive real invoked in Assumption \ref{voisinage}. This means that (\ref{decrease}) holds for  $\bm x^{\bm u^*}_{N}(x_k)$, namely:
\begin{align}
\ell\Bigl(f(\bm x^{\bm u^*}_{N}(x_k),\bar u)\Bigr)&\le \ell\Bigl(\bm x^{\bm u^*}_{N}(x_k))\Bigr)-q(\bm x^{\bm u^*}_{N}(x_k))\nonumber \\
&\le \ell^*_N(x_k)-q(\bm x^{\bm u^*}_{N}(x_k))
\end{align} 
using this last inequality in (\ref{ououfbis}) leads to:
\begin{align}
&J_m(\tilde{\bm u}\vert x_{k+1})\le J^*_m(x_k)-\dfrac{1}{N^m}\ell^*_1(x_k)+\nonumber \\
&(1-\psi(m))\ell^*_N(x_k)-q(\bm x^{\bm u^*}_{N}(x_k))\label{ououfbisbis} 
\end{align} 
which by the definition of $\ell^*_N(x_k):=\ell(\bm x^{\bm u^*}_{N}(x_k))$ and by (\ref{defdegamma}) of Assumption \ref{voisinage} implies: 
\begin{align}
&J_m(\tilde{\bm u}\vert x_{k+1})\le J^*_m(x_k)-\dfrac{1}{N^m}\ell^*_1(x_k)-\nonumber \\
&-(\psi(m)-1+\gamma)\ell^*_N(x_k)\label{ououfbisbis} 
\end{align} 
and therefore, as $\psi(m)\rightarrow 1$ when $m\rightarrow \infty$, there is a finite $m$ beyond which one has:
\begin{align}
&J_m(\tilde{\bm u}\vert x_{k+1})\le J^*_m(x_k)-\dfrac{1}{N^m}\ell^*_1(x_k)-\dfrac{\gamma}{2}\ell^*_N(x_k)\label{ououfbisbisbis} 
\end{align} 
Finally, recalling that $\ell^*_1(x_k)$ is nothing but $\ell(x_{k+1})$ obviously ends the proof. $\hfill \Box$

\begin{rem}[Incorporating state constraints] \label{sconstraints} 
The main changes leading to the incorporation of state constraints are:
\begin{enumerate}
\item The definition of the subset $\mathbb X_N$ should incorporate the constraints satisfaction. \\ 
\item The definition of the local set $B_{\bar\rho}$ should incorporate constraint satisfaction. This guarantee recursive feasibility. 
\end{enumerate} 
Apart from these changes, the proof of the main result is rigorously the same. 
\end{rem}
\section{Conclusion}
In this short note, a new theoretical result is established regarding the stability of a new formulation of finite horizon predictive control incorporating monotonically increasing weight in the stage cost definition. The main result is that the minimal prediction horizon that fits the  target-reachability condition can always be made stabilizing by taking sufficiently high increasing rate for the stage cost penalty. 

\bibliography{biblio_contractive.bib}

\begin{thebibliography}{10}

\bibitem{alamir2007low}
M.~Alamir.
\newblock A low dimensional contractive nmpc scheme for nonlinear systems
  stabilization: Theoretical framework and numerical investigation on
  relatively fast systems.
\newblock In {\em Assessment and Future Directions of Nonlinear Model
  Predictive Control}, pages 523--535. Springer, 2007.

\bibitem{alamir2017}
M.~Alamir.
\newblock Contraction-based nonlinear model predictive control formulation
  without stability-related terminal constraints.
\newblock {\em Automatica}, 75:288--292, 2017.

\bibitem{alamirscl94}
M.~Alamir and G.~Bornard.
\newblock On the stability of receding horizon control of nonlinear
  discrete-time systems.
\newblock {\em Systems \& Control Letters}, 23(4):291 -- 296, 1994.

\bibitem{Alamir1995}
M.~Alamir and G.~Bornard.
\newblock Stability of a truncated infinite constrained receding horizon
  scheme: the general discrete nonlinear case.
\newblock {\em Automatica}, 31(9):1353 -- 1356, 1995.

\bibitem{Boccia:2014}
A.~Boccia, L.~Gr\"{u}ne, and K.~Worthmann.
\newblock Stability and feasibility of state constrained \{MPC\} without
  stabilizing terminal constraints.
\newblock {\em Systems \& Control Letters}, 72:14 -- 21, 2014.

\bibitem{Grim:2005}
G.~Grimm, M.~J. Messina, S.~E. Tuna, and A.~R. Teel.
\newblock Model predictive control: for want of a local control lyapunov
  function, all is not lost.
\newblock {\em IEEE Transactions on Automatic Control}, 50(5):546--558, May
  2005.

\bibitem{Grune2011}
L.~Gr\"{u}ne and J.~Pannek.
\newblock {\em Nonlinear Model Predictive Control. Theory and Algorithms}.
\newblock Springer-Verlag., 2011.

\bibitem{Grune2010}
L.~Gr\"{u}ne, J.~Pannek, M.~Seehafer, and K.~Worthmann.
\newblock Analysis of unconstrained nonlinear mpc schemes with time-varying
  control horizon.
\newblock {\em {SIAM} Journal on Control and Optimization}, 48(8):4938--4962,
  2010.

\bibitem{Jadbabaie:2005}
A.~Jadbabaie and J.~Hauser.
\newblock On the stability of receding horizon control with a general terminal
  cost.
\newblock {\em IEEE Transactions on Automatic Control}, 50(5):674--678, May
  2005.

\bibitem{Keerthi1988}
S.~S. Keerthi and E.~G. Gilbert.
\newblock Optimal infinite horizon feedback laws for a general class of
  constrained discrete-time systems: Stability and moving horizon
  approximations.
\newblock {\em Journal of Optimization Theory and Applications}, 57:265--293,
  1988.

\bibitem{Kothare2000}
S.~Kothare, L.~de~Oliveira, and M.~Morari.
\newblock Contractive model predictive control for constrained nonlinear
  systems.
\newblock {\em IEEE Transations on Automatic Control}, 45, 2000.

\bibitem{Mayne1990}
D.~Q. Mayne and H.~Michalska.
\newblock Receding horizon control of nonlinear systems.
\newblock {\em IEEE Transactions on Automatic Control}, 35:814--824, 1990.

\bibitem{Mayne2000}
D.~Q. Mayne, J.B. Rawlings, C.~V. Rao, and P.~O.~M. Scokaert.
\newblock Constrained model predictive control: Stability and optimality.
\newblock {\em Automatica}, 36:789--814, 2000.

\end{thebibliography}
\bibliographystyle{plain}
\end{document}